\documentclass[conference]{IEEEtran}
\IEEEoverridecommandlockouts
\usepackage{amsmath,amsfonts}
\usepackage{algorithmic}
\usepackage{algorithm}
\usepackage{array}
\usepackage[caption=false,font=normalsize,labelfont=sf,textfont=sf]{subfig}
\usepackage{textcomp}
\usepackage{stfloats}
\usepackage{url}
\usepackage{verbatim}
\usepackage{graphicx}
\usepackage{cite}
\usepackage{comment, float, balance, tabularx, pifont}
\usepackage{tikz}
\usepackage[colorinlistoftodos,textsize=small]{todonotes}
\usepackage{acronym}
\usepackage{booktabs}

\def\BibTeX{{\rm B\kern-.05em{\sc i\kern-.025em b}\kern-.08em
    T\kern-.1667em\lower.7ex\hbox{E}\kern-.125emX}}

\usepackage{scalerel}
\usepackage{tikz}
\usetikzlibrary{svg.path}

\definecolor{orcidlogocol}{HTML}{A6CE39}
\tikzset{
  orcidlogo/.pic={
    \fill[orcidlogocol] svg{M256,128c0,70.7-57.3,128-128,128C57.3,256,0,198.7,0,128C0,57.3,57.3,0,128,0C198.7,0,256,57.3,256,128z};
    \fill[white] svg{M86.3,186.2H70.9V79.1h15.4v48.4V186.2z}
                 svg{M108.9,79.1h41.6c39.6,0,57,28.3,57,53.6c0,27.5-21.5,53.6-56.8,53.6h-41.8V79.1z M124.3,172.4h24.5c34.9,0,42.9-26.5,42.9-39.7c0-21.5-13.7-39.7-43.7-39.7h-23.7V172.4z}
                 svg{M88.7,56.8c0,5.5-4.5,10.1-10.1,10.1c-5.6,0-10.1-4.6-10.1-10.1c0-5.6,4.5-10.1,10.1-10.1C84.2,46.7,88.7,51.3,88.7,56.8z};
  }
}

\newcommand\orcidicon[1]{\href{https://orcid.org/#1}{\mbox{\scalerel*{
\begin{tikzpicture}[yscale=-1,transform shape]
\pic{orcidlogo};
\end{tikzpicture}
}{|}}}}
\usepackage{hyperref}

\newcolumntype{P}[1]{>{\centering\arraybackslash}p{#1}}

\graphicspath{{Figures/}}

\acrodef{DL}{Deep Learning}
\acrodef{RF}{Radio Frequency}
\acrodef{IoT}{Internet of Things}
\acrodef{B5G}{Beyond 5G}
\acrodef{RFF}{Radio Frequency Fingerprinting}
\acrodef{SDR}{Software Defined Radio}
\acrodef{PHY}{Physical}
\acrodef{NLP}{Natural Language Processing}
\acrodef{IoT}{Internet of Things}
\acrodef{SF}{Spreading Factor}
\acrodef{CNN}{Convolutional Neural Network}
\acrodef{DNN}{Dense Neural Network}
\acrodef{LSTM}{Long Short-Term Memory Network}
\acrodef{PCA}{Principal Component Analysis}
\acrodef{t-SNE}{t-Distributed Stochastic Neighbor Embedding }

\begin{document}

 \title{Radio Frequency Fingerprinting via Deep Learning: Challenges and Opportunities}

\author{
    \IEEEauthorblockN{Saeif Al-Hazbi~\orcidicon{0000-0002-7884-5025}\IEEEauthorrefmark{1}, Ahmed Hussain~\orcidicon{0000-0003-4732-9543}\IEEEauthorrefmark{2}, Savio Sciancalepore~\orcidicon{0000-0003-0974-3639}\IEEEauthorrefmark{3}, Gabriele Oligeri~\orcidicon{0000-0002-9637-0430}\IEEEauthorrefmark{1}, Panos Papadimitratos~\orcidicon{0000-0002-3267-5374}\IEEEauthorrefmark{2}}
    \IEEEauthorblockA{
    \IEEEauthorrefmark{1}College of Science and Engineering (CSE), Hamad Bin Khalifa University (HBKU) -- Doha, Qatar
    \\\{salhazbi, goligeri\}@hbku.edu.qa}
    \IEEEauthorblockA{\IEEEauthorrefmark{2}Networked Systems Security group, KTH Royal Institute of Technology -- Stockholm, Sweden
    \\ahmed.hussain@ieee.org, papadim@kth.se}
    \IEEEauthorblockA{\IEEEauthorrefmark{3}Eindhoven University of Technology -- Eindhoven, Netherlands. s.sciancalepore@tue.nl}
}

\maketitle

\let\svthefootnote\thefootnote
\newcommand\freefootnote[1]{%
  \let\thefootnote\relax%
  \footnotetext{#1}%
  \let\thefootnote\svthefootnote%
}

\begingroup\renewcommand\thefootnote{\textsection}\freefootnote{This is a personal copy of the authors. Not for redistribution. The final version of the paper will be available in the IWCMC 2024 Conference Proceedings and IEEE Xplore.}

\begin{abstract}
\acf{RFF} techniques promise to authenticate wireless devices at the physical layer based on inherent hardware imperfections introduced during manufacturing. Such RF transmitter imperfections are reflected into over-the-air signals, allowing receivers to accurately identify the RF transmitting source. Recent advances in Machine Learning, particularly in \acf{DL}, have improved the ability of RFF systems to extract and learn complex features that make up the device-specific fingerprint. However, integrating \ac{DL} techniques with \ac{RFF} and operating the system in real-world scenarios presents numerous challenges, originating from the embedded systems and the \ac{DL} research domains. This paper systematically identifies and analyzes the essential considerations and challenges encountered in the creation of \ac{DL}-based RFF systems across their typical development life-cycle, which include (i) data collection and preprocessing, (ii) training, and finally, (iii) deployment. Our investigation provides a comprehensive overview of the current open problems that prevent real deployment of \ac{DL}-based \ac{RFF} systems while also discussing promising research opportunities to enhance the overall accuracy, robustness, and privacy of these systems.

\end{abstract}


\section{Introduction}
\label{sec:intro}

\acf{RFF} techniques have recently gained attention in the scientific community as a way to authenticate \ac{RF} devices based on samples of their \ac{PHY} layer signals~\cite{soltanieh2020review}. \ac{RFF} grounds on the assumption that two RF devices are unlikely to transmit signals with the same \ac{PHY} layer features, even when they convey the same data. Therefore, a rogue transmitter that tries to mimic a legitimate \ac{IoT} device by replaying its messages would inevitably leave its own fingerprint, distinct at the \ac{PHY} layer from that of the legitimate device.
This unique property can serve as either an alternative or a direct complement to existing cryptography-based authentication protocols. In particular, being orthogonal to cryptographic protocols and requiring no modifications to the transmitter's hardware or software, \ac{RFF} is well-suited for situations where updates are difficult (e.g., satellite networks) or devices are characterized by strict hardware constraints (e.g., \ac{IoT})~\cite{oligeri2023}.

\begin{figure}
    \centering
    \includegraphics[width=\columnwidth]{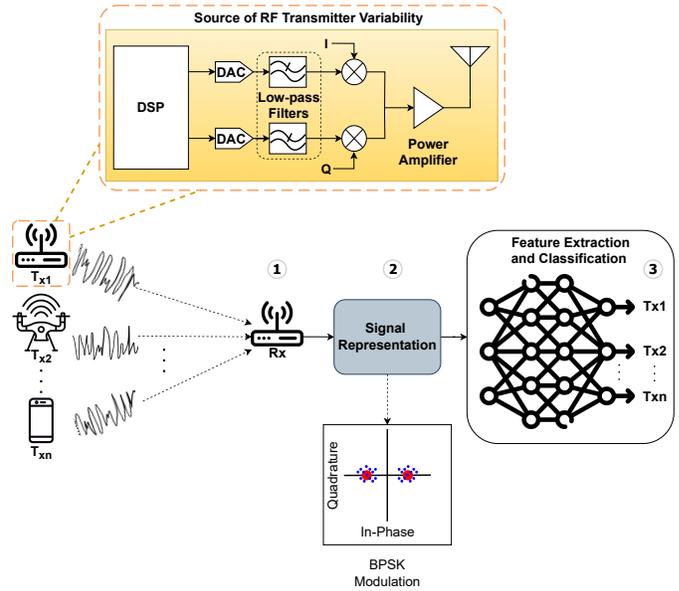}
    \caption{\ac{RFF} building blocks: The receiver identifies the transmitting source by analyzing the received signal at the PHY layer.}
    \label{fig:RFF_Concept}
\end{figure}

The identification and extraction of \ac{PHY} layer features required for RFF involve integrating advanced data mining techniques as part of the system. Specifically, the use of \acf{DL} techniques in \ac{RFF} systems has shown significant potential to improve performance, due to the remarkable ability of \ac{DL} models to identify complex patterns and extract relevant features from raw data~\cite{lecun2015deep}. The effectiveness of \ac{DL}-based \ac{RFF} systems in accurately identifying \ac{RF} devices has been reported in the literature, but these outstanding results were obtained primarily under controlled laboratory conditions. At the time of this writing, \ac{RFF} deployment remains challenging in real-world applications.

Factors such as the calibration and configuration of the \ac{DL} methods, especially within the context of \ac{RFF} systems, the unpredictability of the radio channel, and the deployment conditions contribute to the difficulties of adapting \ac{DL} solutions for \ac{RFF} systems. Challenges associated with \ac{DL}-based \ac{RFF} systems are discussed in several recent studies~\cite{jagannath2022_comnet, hanna2022wisig, rehman2014radio, elmaghbub2021comprehensive, 9580899}; however, such considerations are disconnected from the real-world deployment of such systems.

Thus, we currently miss a systematic classification of the main challenges that prevent the adoption of \ac{DL}-based \ac{RFF} systems on a large scale in real-world scenarios.

{\bf Contribution.} In this paper, we systematically identify and classify the building blocks necessary for assembling \ac{DL}-based \ac{RFF} systems, as well as the challenges that currently prevent their real-world deployment. 

We do so with reference to the standard development pipeline of \ac{DL}-based \ac{RFF} systems, which includes: \emph{RF Data Collection and Preprocessing}, \emph{Training}, and \emph{Deployment}. 

For all identified challenges and each phase, we provide explanations and examples grounded in the literature. Addressing such challenges can be a springboard for \ac{DL}-based \ac{RFF} deployment in real-world scenarios and further future research. Essentially, we bridge the gap between research and practice towards \ac{DL}-based \ac{RFF} deployment.

Our contribution is not yet another survey on \ac{RFF}. This paper addresses researchers in Industry and Academia approaching the \ac{RFF} domain, providing a tutorial-style discussion of the aspects connected to both the \ac{DL} and the embedded systems areas and their intersection.

The remainder of this paper is organized as follows. Section~\ref{sec:background} introduces \ac{DL}-based \ac{RFF} systems.    
Section~\ref{sec:pre-training} highlights key considerations required for both the stage of data collection and preprocessing as well as the training, and Section~\ref{sec:deployment} discusses deployment challenges. We highlight future research opportunities in Section~\ref{sec:future}, before we conclude in Section~\ref{sec:conclusion}.

\begin{figure*}  
    \centering    
    \includegraphics[width=\textwidth]{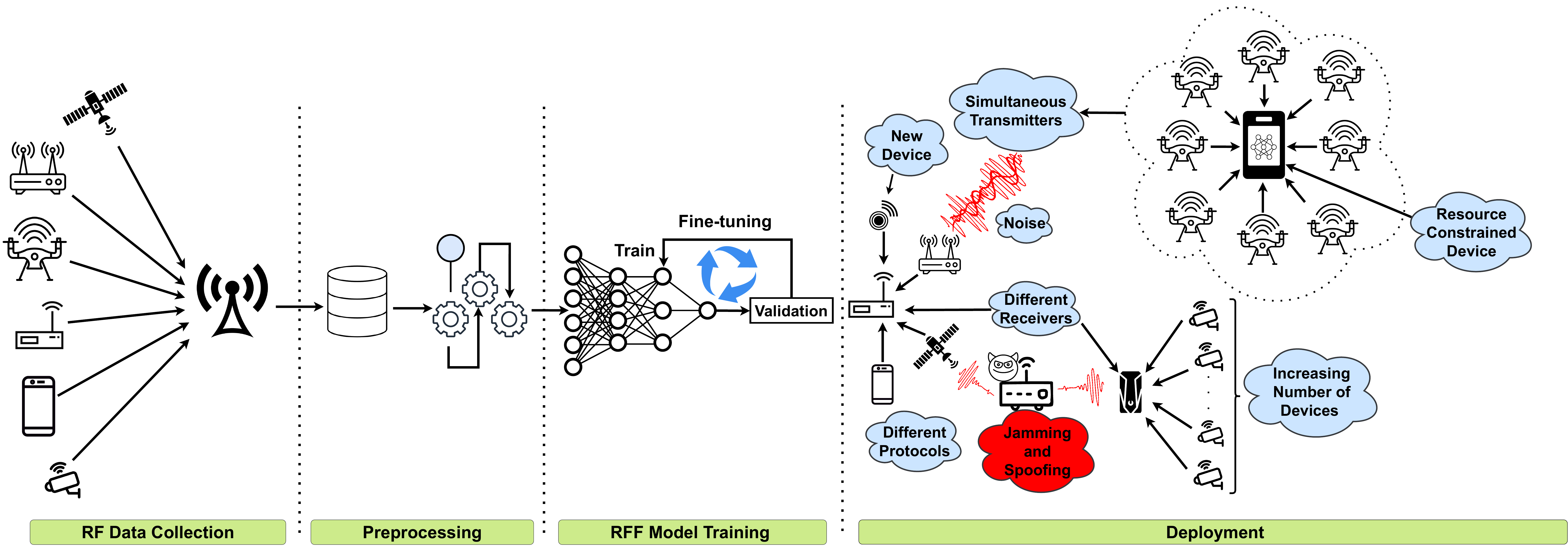}
    \caption{Typical workflow of \ac{DL}-based \ac{RFF}.}
    \label{fig:workflow}
\end{figure*}

\section{Preliminaries on DL-based RFF}
\label{sec:background}

\ac{RFF} techniques identify wireless devices based on their unique signal characteristics. As shown in Fig.~\ref{fig:RFF_Concept}, \ac{RF} devices emit unique waveforms in the radio spectrum due to manufacturing imperfections in various components, e.g., digital-to-analog converters, mixers, power amplifiers, antennas, etc. Once captured by a receiver, these signals can be represented in various formats and then processed by a \ac{DL} model to identify and extract subtle variations per device. These imperceptible variations inherent to each signal constitute the signature of the device and serve as the foundation for authentication. The process of designing and implementing \ac{DL}-based \ac{RFF} systems typically involves three essential phases, known as the \ac{DL} pipeline: (i) {\em data collection and pre-processing}, (ii) {\em training}, and finally (iii) {\em deployment} (see Fig.~\ref{fig:workflow}). 

{\bf Data collection and Pre-processing.} \ac{RF} signals emitted by wireless devices within a monitored environment are captured and stored as In-phase (I) and Quadrature (Q) samples. This format preserves information about the amplitude and phase distortions. After acquisition, the acquired data are prepared for model training through a series of preprocessing techniques, including, e.g., feature selection, data augmentation, normalization, and noise removal. In the context of the \ac{RFF} building pipeline, data preprocessing is particularly critical given that data are collected from the wireless channel, where \ac{RF} signals might experience significant unpredictable and time-varying distortion due to phenomena such as multipath propagation, interference, shadowing, and fading.

{\bf Training.} This phase involves designing and training a \ac{DL} architecture capable of effectively capturing the underlying patterns corresponding to the unique \ac{RF} fingerprints of wireless devices. Factors such as architecture selection, hyper-parameter tuning, processing variable input lengths, and designing proper validation and testing metrics are crucial for building robust and accurate \ac{RFF} systems. Throughout this phase, the model undergoes iterative fine-tuning and rigorous evaluation to ensure accuracy, generalization, and adaptability.

{\bf Deployment.} The trained model is integrated into a production system to perform the authentication of wireless devices. In this context, authentication is achieved by comparing incoming RF fingerprints with those learned during training. The \ac{RFF} system should efficiently and reliably authenticate wireless devices under various environmental conditions and deployment scenarios. However, various issues, such as system scalability, adaptability to new conditions, and robustness to adversarial attacks, affect the system's performance.

\section{Data Collection, Pre-processing, and Training}
\label{sec:pre-training}
\textbf{Training Dataset Collection Settings.} Existing techniques to collect \ac{RF} fingerprints typically involve generating short, intermittent snapshots of device wireless data over several days~\cite{elmaghbub2021comprehensive,al2020exposing}. The aim is to capture the \ac{RF} fingerprint across varying wireless channel conditions and ensure a manageable dataset size for efficient training. However, these techniques often overlook additional factors that could impact the devices' actual \ac{RF} fingerprint, including effects from power cycles, temperature variations in internal circuits, and device aging~\cite{saeif2023day}. In real-world scenarios, \ac{RF} devices operate continuously, posing a challenge in developing data collection methods that can effectively capture the dynamic nature of the device fingerprint. 

\textbf{Input Data Selection.} Each segment of the transmitted wireless signal possesses unique characteristics, potentially useful for \ac{RFF}. However, selecting the optimal data segment as input to the \ac{RFF} model can be challenging. Ideally, the chosen data segment should show a consistent and repetitive pattern. In fact, this can guarantee that the \ac{DL} model learns from the unique patterns of the RF signal rather than becoming biased towards the specific content of the wireless segment. The preamble of the wireless packet, containing synchronization-related information, is particularly appealing as it is consistent nature across different devices and packets (for the same communication technology). Overall, selecting the optimal wireless signal segment for \ac{RFF} is still an open challenge that requires further research and investigation.

\textbf{Data Augmentation.} Refers to a set of established transformations applied to the original data to diversify it while preserving its statistical properties. The main objective is to increase the size of the data set and improve the generalization and robustness of the model. For example, in computer vision, augmentation techniques such as flipping, rotation, scaling, and cropping are used successfully to expand the data and mitigate overfitting. However, in the context of \ac{RFF}, identifying universally effective techniques to enhance wireless data remains challenging. 

This issue arises due to the complexity and dynamic nature of wireless signals, which are subject to channel conditions and hardware-based effects.

\textbf{Signal Representation.} Wireless signals can be represented in various forms using different signal processing methods, each capturing different characteristics of the signal. The features extracted from these analyses significantly impact the performance of the \ac{DL} model. Representations such as raw IQ data, spectral characteristics, and the combination of time- and frequency-related metrics are often considered~\cite{jagannath2022_comnet}. Raw IQ data capture amplitude and phase variations caused by hardware imperfections. Spectral analysis extracts frequency-domain characteristics of the \ac{RF} signal. Furthermore, the time-frequency representations capture the combined time-varying frequency behavior of the \ac{RF} signal. Since each representation has its own strengths and weaknesses, selecting the most suitable one for training \ac{RFF} models is not straightforward and requires further research. 

\textbf{Feature Normalization.} Wireless data is inherently dynamic and subject to time-varying channel conditions, hardware imperfections, and noise, all of which can introduce power fluctuations and scaling issues in the received data. Given that \ac{DL} models are scale-sensitive, these inconsistencies, if not addressed, can negatively affect the models' learning and generalization. For example, when training with raw IQ data, significant scale differences between the I and Q components may lead the model to assign greater weight to the component with a larger scale, resulting in a scale-dependent solution for device identification. By equalizing all wireless data feature ranges (normalization), \ac{DL} models can correctly capture the underlying pattern of the data distribution and reduce potential biases towards signal power or feature scale.

\textbf{Neural Network Structure \& Hyper-parameter Tuning.} The \ac{DL} architecture is crucial for the overall accuracy and robustness of \ac{RFF} systems. Many available solutions reuse or modify existing \ac{DL} networks, reporting outstanding performance in other domains, such as computer vision and \ac{NLP}~\cite{jagannath2022_comnet}.  However, these models, initially designed to capture domain-specific features, cannot be expected to perform well with wireless data. Therefore, it becomes imperative to develop specialized \ac{DL} models to deal with the complexity of wireless data and adapt to its inherently dynamic nature. Fig.~\ref{fig:fixediq_randomiq} compares neural network architectures trained on preamble raw IQ data with a defined sequence. To assess model robustness against temporal changes, the sequence of the preamble data is randomized during testing. While all models achieved high training accuracy, only the Transformer architecture (\emph{without positional encoder}) demonstrated strong generalization on both training and testing data, regardless of the raw IQ data sequence. Furthermore, the configuration of \ac{DL} models with different hyper-parameter values can affect their performance during training and deployment. Suboptimal hyper-parameter selection might cause overfitting or underfitting, resulting in degraded performance. For example, analyzing the impact of batch size and epochs, Fig.~\ref{fig:batchsize_variation} illustrates how varying the batch size affects the model learning. Smaller batch sizes achieve higher accuracy and converge faster but are more computationally demanding, whereas larger batches converge more slowly but demonstrate steady progress over more epochs.

\begin{figure*}[ht!]
\centering
\subfloat[\label{1a}]{
    \includegraphics[width=0.3\textwidth]{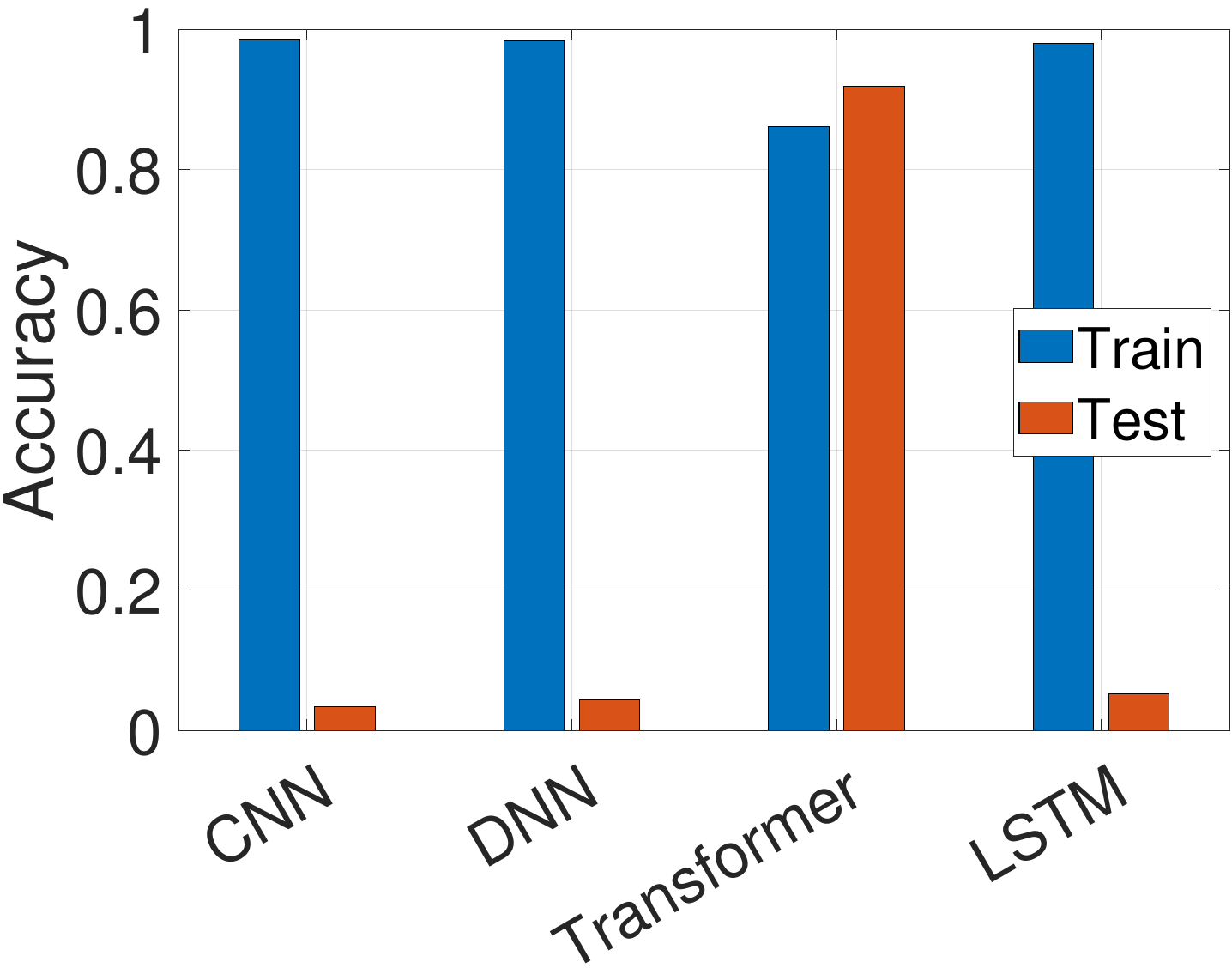}
    \label{fig:fixediq_randomiq}
}\subfloat[\label{1b}]{
    \includegraphics[width=0.3\textwidth]{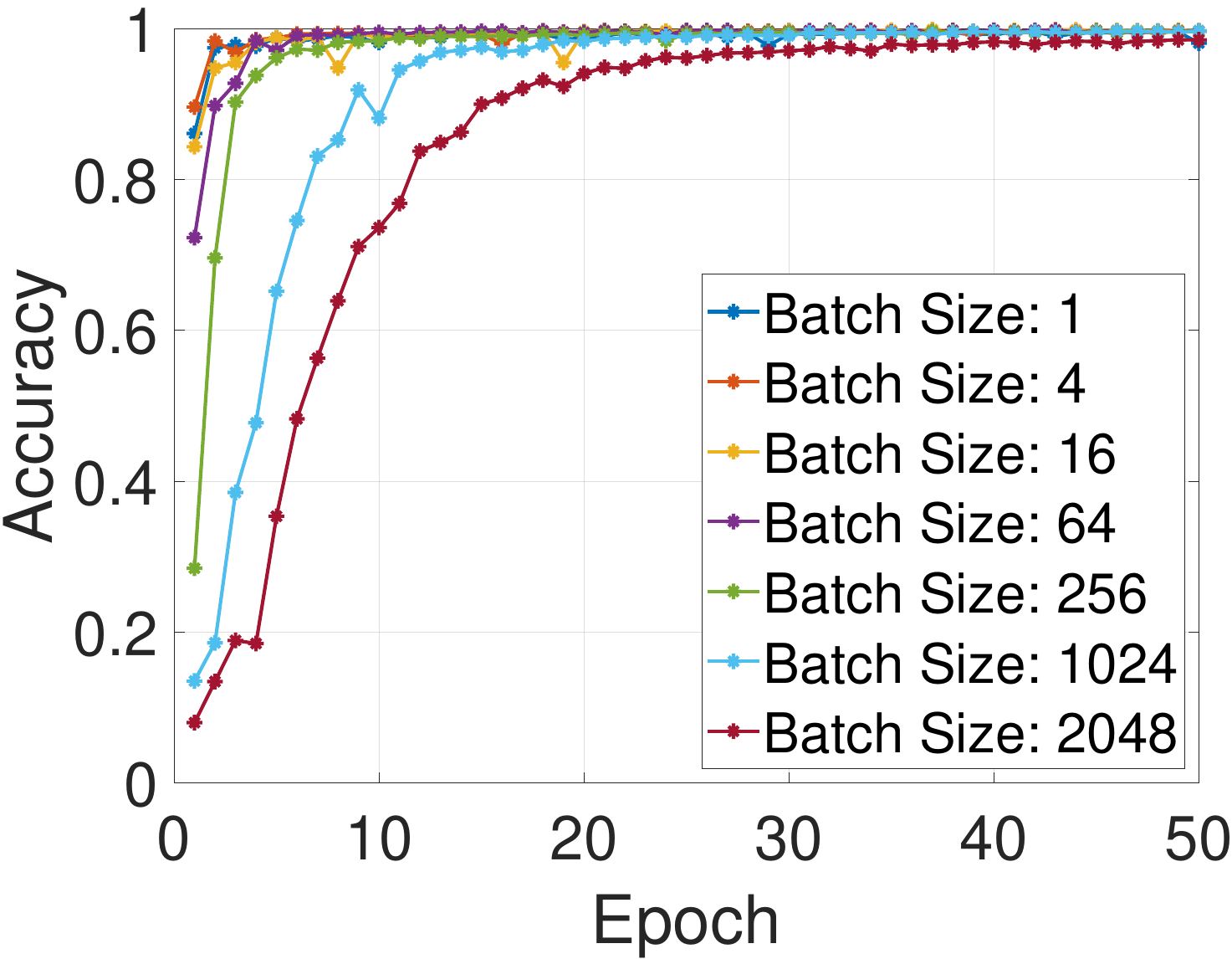}
    \label{fig:batchsize_variation}
}

\caption{Impact of \ac{DL} architecture and hyper-parameter tuning on model performance based on the analysis of the dataset released in~\cite{hanna2022wisig}. (a) Comparison of different \ac{DL} architectures, including \acfp{CNN}, \acfp{DNN}, Transformers, and \acfp{LSTM}, trained on fixed, consistent raw IQ sequences. The tests were performed on randomly shuffled raw IQ sequences. (b) Validation accuracy as a function of the number of epochs for various batch sizes.}
\label{fig:performance_based_on_order_and_content}
\end{figure*}

\textbf{Design of a Proper Validation/Testing Methodology.} While conventional validation techniques like \emph{Holdout} or \emph{K-fold cross-validation} can be utilized to evaluate \ac{DL} models with wireless data, the dynamic nature of wireless data requires special considerations. Wireless channels frequently experience time-varying conditions, e.g., fading, multipath propagation, and Doppler effects. These elements change the data distribution over time, making it difficult to guarantee that the validation and testing sets are representative of future data~\cite{al2020exposing}. Moreover, spatial variability introduces another level of complexity, as specific physical environments can greatly influence wireless data. Consequently, it is imperative to validate and test the model on diverse data captured from different locations and periods, as this will provide an accurate assessment of its performance under real-world conditions.

\begin{figure*}
    \centering
  \subfloat[\label{1a_cos_sim}]{%
       \includegraphics[width=0.3\textwidth]{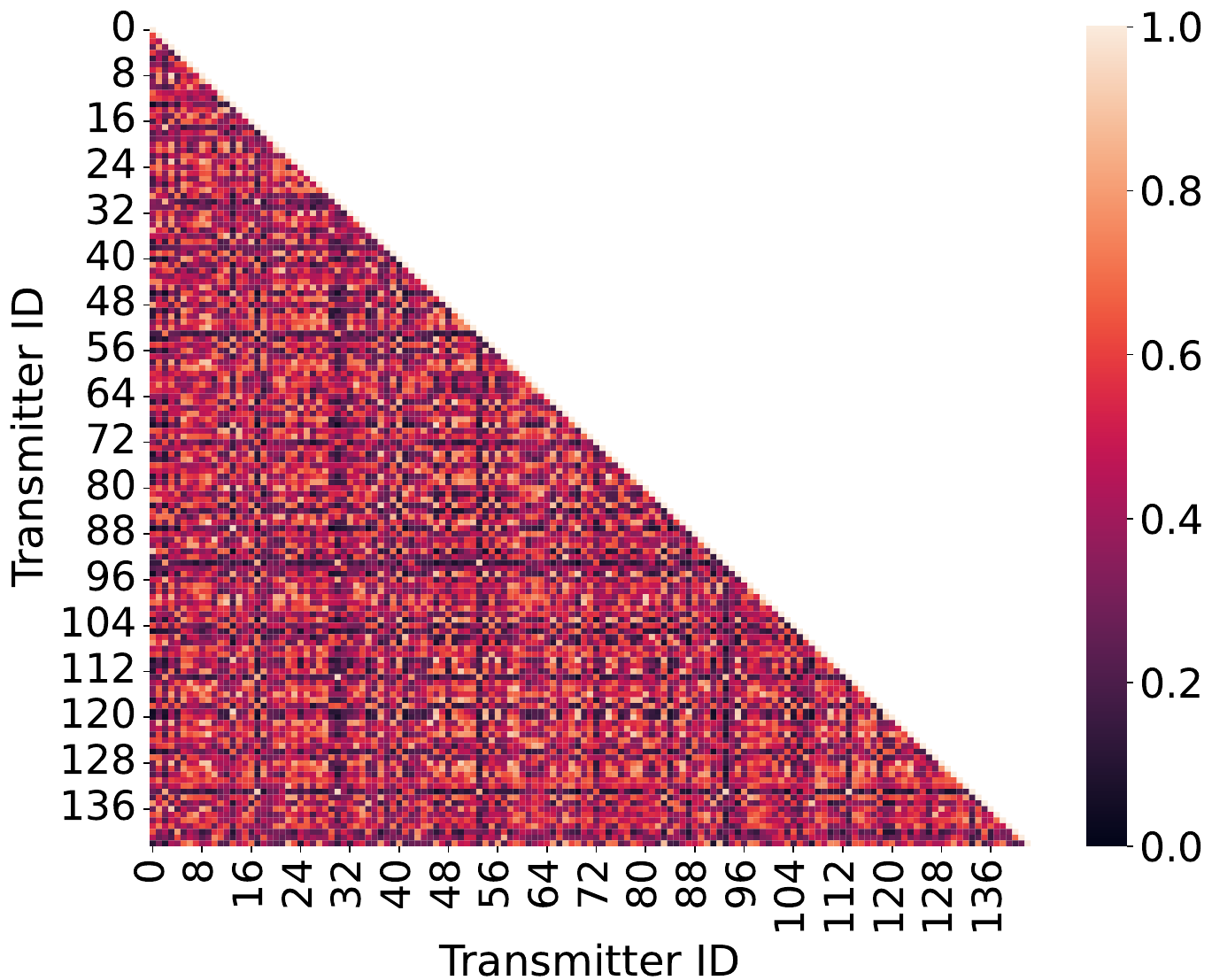}}
  \subfloat[\label{2b_pca}]{%
        \includegraphics[width=0.3\textwidth]{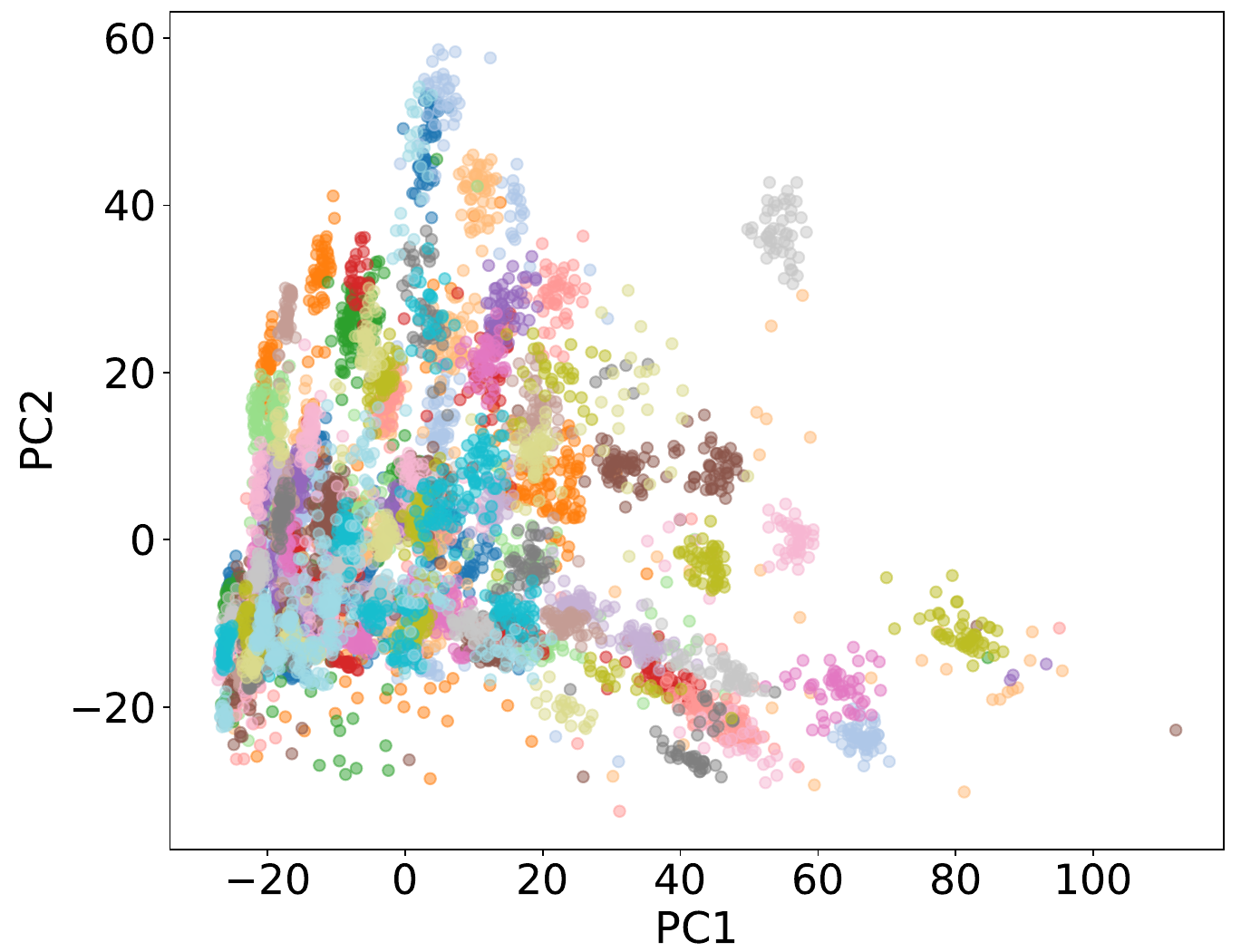}}
  \subfloat[\label{3c_tsne}]{%
        \includegraphics[width=0.3\textwidth]{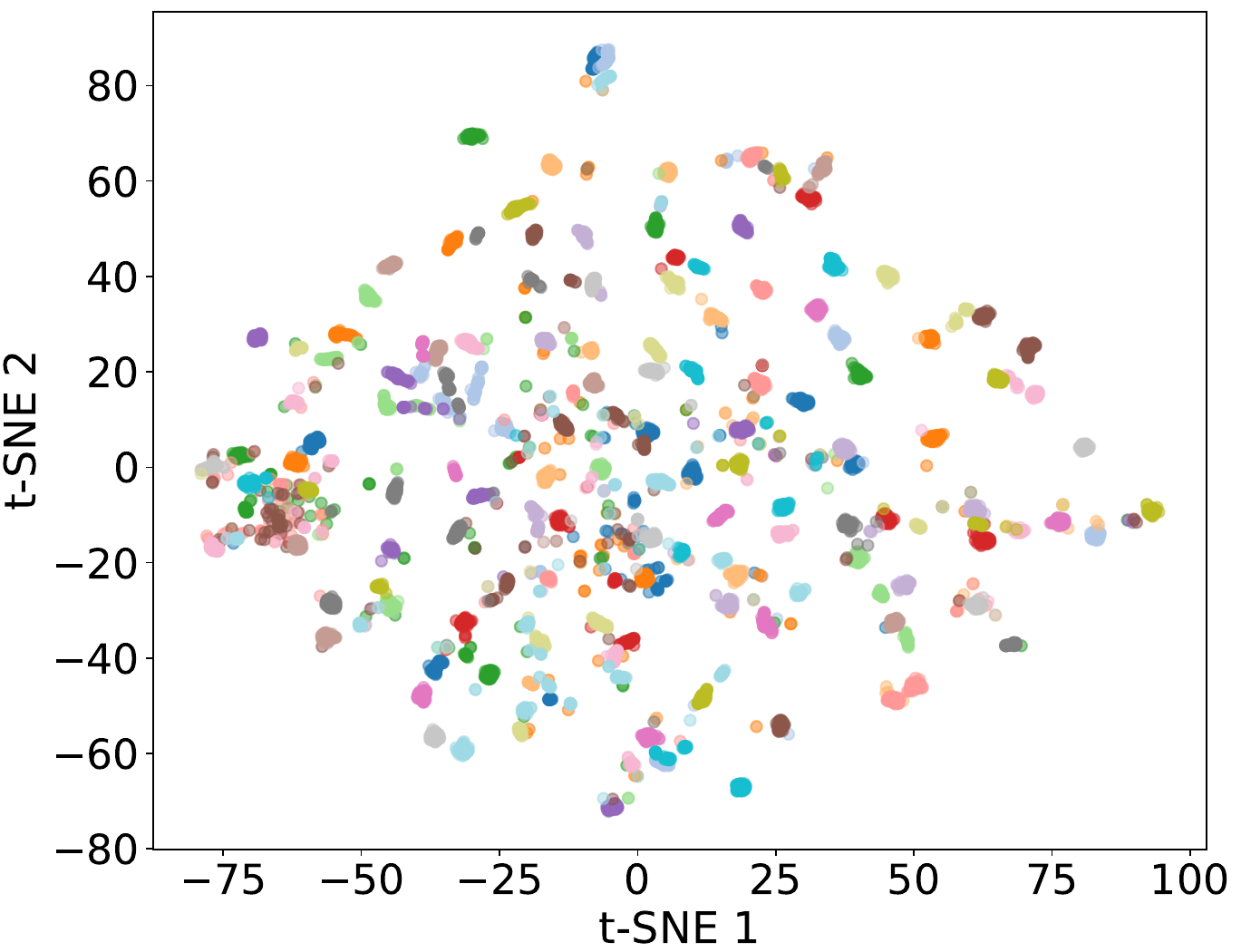}}
  \caption{Visualization of feature embedding for a reference \ac{DL}-based \ac{RFF} model trained to identify 143 transmitters, based on our own analysis of the dataset released in~\cite{hanna2022wisig}. (a) Cosine similarity analysis highlights the degree of similarity among features. (b) Projection using \acf{PCA} shows the distribution variation and clustering of similar transmitters. (c) \acf{t-SNE} visualization of the transmitters' features, providing a dimensionality-reduced view of the crowded feature space and its overlapping regions. The colors in (b) and (c) represent individual transmitters. 
  }
  \label{fig:feature_visualization}
\end{figure*}

\section{Deployment Challenges}
\label{sec:deployment}
\textbf{Model Interpretability Issues.} Although \ac{DL} techniques often show remarkable performance compared to other Machine Learning techniques, their \emph{blackbox} nature and lack of interpretability are significant drawbacks. In particular, it is difficult to explain how these models classify the data and which specific underlying features contribute to the final decision produced by the model. This issue becomes even more relevant when considering that \ac{DL}-based \ac{RFF} systems are deployed for security applications, such as device authentication and intrusion detection. Given the security implications, a transparent understanding of the rationale of the decision-making process of these models is even more critical to mitigate the risk of false predictions while maintaining robustness and reliability.

\textbf{Lack of Benchmark Dataset.} Standardized benchmark datasets for computer vision and \ac{NLP} provide a common ground for comparing the performance of \ac{DL} models in those domains. On the contrary, for \ac{RFF}, developing a widely accepted benchmark dataset remains a challenge. To overcome this, researchers often resort to locally generated datasets for model validation. Although such datasets could be practical for initial model testing, they may only represent a subset of real deployment conditions. 

This bias may affect the robustness and generalizability of the model when deployed in real-world scenarios. In general, creating a comprehensive and standardized RFF benchmark dataset is crucial to evaluate deployed \ac{DL}-based \ac{RFF} models across various modulation schemes, transmission rates, communication technologies, and channel conditions.

\textbf{Receiver Hardware Bias.} Most existing \ac{RFF} solutions collect data using a single receiver to identify a specific wireless transmitter within a pool of devices. Using this technique, it is commonly assumed that the receiver hardware does not introduce its own variability to the captured \ac{RF} fingerprint. Additionally, the receiver hardware used during training is typically expected to remain unchanged during deployment. In reality, the RF fingerprint is affected by the entire communication chain, including the transmitter, the channel, and finally, the receiver~\cite{merchant2019toward,al2020exposing}. Therefore, the resulting trained model is significantly biased towards the physical characteristics of the receiver used during training. With this limitation, any change at the receiver---where the model has been deployed---would require retraining the model.
This is paramount for any generalized deployment, with a trained model for one transmitting device, $\mathcal{V}$, distributed to many (or any) devices that seek to identify/authenticate $\mathcal{V}$.

\textbf{Scalability Issues.} \ac{DL}-based \ac{RFF} systems are designed to be deployed in large-scale network environments with numerous devices. However, as the number of devices increases, the performance of the \ac{DL}-based \ac{RFF} might degrade~\cite{hanna2022wisig}. This degradation is mainly due to the difficulty in distinguishing devices with similar characteristics. For instance, in a network employing BPSK modulation, using raw IQ data for device authentication could introduce scalability issues. Specifically, the presence of limited features (strong I and nearly negligible Q) combined with a densely populated feature space can lead to overlapping features from similar devices, resulting in performance degradation. Fig.~\ref{fig:feature_visualization} presents a visualization of the feature embedding and cosine similarity for a model designed to identify 143 wireless transmitters when analyzing the dataset in~\cite{hanna2022wisig}. As shown in Fig.~\ref{1a_cos_sim}, cosine similarity reveals areas of high similarity between the transmitter features. Moreover,  Fig.~\ref{2b_pca} and~\ref{3c_tsne} reveal a dense feature space and many overlapping regions, indicating the potential challenge for the model to distinguish transmitters. Overall, improving the scalability of \ac{DL} based \ac{RFF} systems requires developing novel solutions, which may involve improved feature engineering and model architecture design.

\textbf{Protocol Dependency.} Current \ac{DL}-based \ac{RFF} systems are inherently protocol-dependent and sensitive to protocol configuration, which in turn affects their device identification capabilities~\cite{9580899}. For example, WiFi and LoRa define a range of parameters, including modulation schemes, bit rates, spreading sequences (in the case of LoRa), allowed frequencies, bandwidth, and transmit power, which can be fine-tuned based on operational conditions. Changes in these parameters affect the acquired RF fingerprint, leading to a decrease in overall system performance. This performance drop is especially noticeable when training and testing data are collected under different protocol configurations.

\textbf{Detection of Unauthorized Devices.} Most existing \ac{DL}-based \ac{RFF} systems operate under closed-set classification conditions, i.e., they only recognize devices that were included in their training set. While this approach works well in controlled environments, it presents a significant challenge when new, previously unseen devices transmit data.
In such a scenario, closed-set \ac{DL}-based \ac{RFF} systems cannot reliably recognize new devices, primarily because they have not encountered their unique RF fingerprints during the training phase. As a result, the system compares the new unseen devices based on the one(s) that closely match the \ac{RF} fingerprint from its training set. To overcome this challenge, it is crucial to develop open-set classification techniques (e.g., zero-shot learning) that can identify, handle, and reject new unauthorized devices during the deployment phase, thus improving the security and adaptability of the system.

\textbf{Simultaneous Transmissions.} During training, \ac{RF} profiles of transmitting devices are usually acquired by isolating a specific transmitter device that operates on a single frequency channel and collecting data with a receiver~\cite{gutierrez2022considerations}. However, in real-world \ac{RFF} environments, multiple devices may transmit simultaneously, either on the same frequency or across different channels. Such concurrent transmissions can degrade the performance of \ac{DL}-based \ac{RFF} systems due to signal overlap at the receiver. Furthermore, any change in the training channel profile can further affect performance, as the system may encounter unfamiliar channel characteristics.

\textbf{Deployment on Constrained Devices.} Most current \ac{DL}-based \ac{RFF} studies focus on developing \ac{RFF} systems intended to be deployed in computing environments with ample processing power, memory, and energy. However, in many real-world scenarios, deploying these computationally intensive models on lightweight devices is often impractical. This is particularly relevant for scenarios such as edge computing and resource-constrained IoT devices. Therefore, more research is required to explore optimization techniques that allow \ac{RFF} models to run efficiently on lightweight devices without compromising performance.

\textbf{Variable Input Length.} Wireless signals can vary in size and duration, depending on the specific communication technology and modulation scheme~\cite{10100932}. This variability is particularly evident in adaptive communication protocols such as LoRaWAN, where the length of transmitted symbols varies in response to network conditions and proximity. This is challenging for conventional \ac{DL} models, designed to handle fixed-size input data. These models require the input data to be truncated or padded to fit the input size, which can potentially lead to information loss and increase the computational resources to compensate for missing data. However, the most critical implications lie in the negative impact on overall model performance. 

\textbf{System Adaptiveness.} Existing \ac{DL}-based \ac{RFF} systems typically operate under the assumption that the set of authorized devices present during the training phase remains the same at the time of the deployment phase. However, real-world wireless networks are often dynamic, and devices frequently join and leave the network. The current solution to this challenge involves retraining the system model whenever there is a change in the network---a computationally expensive and time-consuming process. This issue becomes particularly critical in large critical infrastructures, where updates must be executed seamlessly without disrupting regular operations.

\textbf{Inference Time.} In real-world settings, the time to perform the tests, namely the inference time, is a key factor in determining the suitability of \ac{DL}-based \ac{RFF} systems to achieve their objective. Short inference times are particularly important in time-sensitive applications, where delays cannot be tolerated. To ensure the practicality of \ac{DL}-based \ac{RFF} systems, it is essential to evaluate and optimize the model inference time across a range of devices that could potentially be deployed within the specific network. This evaluation and optimization is necessary, given that the inference time of \ac{DL} models varies based on the hardware components in which they are installed. 

\textbf{Robustness to Channel Variations.} \ac{DL}-based \ac{RFF} systems rely on data collected from the physical layer of the wireless spectrum. However, these data are subject to strong channel fluctuations, noise, multipath, and shadowing effects caused by environmental changes. These channel variations can significantly affect the resulting profile of \ac{RF} signals, making the channel conditions experienced during training often different from those experienced during the deployment phase. Eventually, this variability leads to performance degradation~\cite{al2020exposing}.

\textbf{Jamming.} Jamming attacks occur when a malicious entity intentionally transmits a signal on the same frequency band as the target communication link, disrupting ongoing communication. In our context, jamming affects the reception of the clean \ac{RF} signal, making the extraction of \ac{RFF} features challenging. As mentioned above, this challenge is exacerbated as the \ac{RF} signal is already affected by other noise sources present in the wireless channel. As a result, jamming attacks can significantly compromise the reliability of the system.

\textbf{Spoofing.} The openness of wireless communications allows malicious entities to eavesdrop on radio signals and inject new signals to gain unauthorized access. In the \ac{RFF} context, an attacker can capture the unique \ac{RF} fingerprint of a target device, emulate or distort the signal to create a deceptive one, and then transmit it to try to impersonate the legitimate device. 
More research is needed to validate the possibility of this adversarial scenario and test the resilience of \ac{DL}-based \ac{RFF} systems against this attack.

\textbf{Adversarial Attacks.} \ac{DL} models are  vulnerable to adversarial attacks~\cite{goodfellow2014explaining}. For example, in image classification tasks, malicious entities can trick the model by manipulating the input imperceptibly to the human eye, but can fool the model into making wrong predictions. In the context of \ac{RFF}, adversarial attacks can significantly impact the performance and reliability of \ac{DL}-based \ac{RFF} systems.
By exploiting the sensitivity of the model to input perturbations, an attacker can carefully craft deceptive adversarial samples to compromise the integrity of the system.
At the time of this writing, only a limited number of studies have investigated the robustness of \ac{DL}-based \ac{RFF} systems to such attacks, and only under ideal conditions, without considering the requirements necessary for executing successful attacks.

\section{Research Opportunities}
\label{sec:future}
\textbf{Realistic Channel Simulation.} One promising approach to mitigate channel variability involves simulating \ac{RF} fingerprint data under various realistic wireless channel conditions, with data taken from real data wireless channels~\cite{yan2022rrf}. Although this research direction is still in its early stages, it holds significant potential, as it exposes the signal to real-world deployment conditions and introduces variability and augmentation to the data before training.

\textbf{Continual Learning.} Another effective strategy might involve integrating continuous adaptive learning into their deployment. By adopting this approach, these systems can continuously update in real-time without being re-trained, allowing them to maintain accuracy and adapt to the highly dynamic nature of wireless channels while being deployed.

\textbf{Collaborative Learning and Multimodal Approach.} To improve the reliability and accuracy of \ac{DL}-based \ac{RFF} systems, researchers can investigate the concept of multimodal and collaborative learning where different types of input data or model predictions are fused together to strengthen the overall system's performance.

\textbf{\ac{RFF} Anonymization.} Once the \ac{RF} device fingerprint is leaked, it can be used against the targeted device. Specifically, malicious entities can exploit it for activities such as device tracking, profiling, and spoofing. Consequently, developing privacy-preserving \ac{RFF} techniques is crucial to protect individual privacy and data while ensuring the usability of these \ac{DL}-based \ac{RFF} systems.

\section{Conclusion}
\label{sec:conclusion}
Research on reliable and robust \ac{DL}-based \ac{RFF} systems has progressed significantly in recent years, yielding effective solutions for various scenarios and applications. However, deploying \ac{RFF} systems in real-world settings still faces numerous challenges inherent to \ac{DL} techniques and the highly dynamic nature of wireless scenarios. This paper has systematically identified and analyzed the most significant considerations and challenges towards deploying real-world \ac{DL}-based \ac{RFF} systems, categorized per phase in the development pipeline. We also outlined promising future research opportunities. The aim is to provide researchers and industry experts with succinct guidance on the main challenges and assist efforts toward real-world \ac{RFF} deployment.

\section*{Acknowledgement}
This research was made possible by the award GSRA7-1-0510-20045 and NPRP12C-0814-190012-SP165 from Qatar National Research Fund (a member of Qatar Foundation) and by the INTERSECT project, Grant ID NWA.1162.18.301, funded by Netherlands Organisation for Scientific Research (NWO). The work of P. Papadimitratos and A. Hussain was supported in part by the Swedish Science Foundation (VR). The contents herein are solely the responsibility of the author(s). 

\bibliographystyle{IEEEtran}
\balance
\bibliography{main}

\end{document}